\documentstyle[12pt]{article}    
           
        \newcommand{\be}{\begin{equation}}
        \newcommand{\ee}{\end{equation}}
        \newcommand{\bea}{\begin{eqnarray}}
        \newcommand{\eea}{\end{eqnarray}}
        \newcommand{\za}{\langle  }
        \newcommand{\ze}{ \rangle}
        \def \3{\ss}

\begin{document}

\begin{center}
  {\Large \bf        On the Local Equilibrium Condition }
\end{center}
    
\vspace{10mm}

\begin{center}
                     Hermann He\3ling 
\end{center}

\begin{center}
                     FHTW Berlin \\
                     University of Applied Sciences \\
                     Treskowallee 8 \\
                     D--10318 Berlin \\
                     Germany
\end{center}

\vspace*{1mm}

\begin{center}
\end{center}

\vspace*{1mm}

\begin{abstract}

A  physical system should be in a local equilibrium if it cannot be
distinguished from a global equilibrium by ``infinitesimally localized
measurements''.
This seems to be a natural characterization of local equilibrium, however
the problem is to give a precise meaning to the qualitative phrase
``infinitesimally localized measurements''. 

A solution is suggested in form of a 
{\em Local Equilibrium Condition} (LEC) which
can be applied to non-interacting scalar quanta.

The Unruh temperature of massless quanta is 
derived by applying LEC to an arbitrary point 
inside the Rindler Wedge.

Massless quanta outside a hot sphere are analyzed.
A stationary spherically symmetric 
local equilibrium does only exist according to LEC
if the temperature is globally constant.

Using LEC a non-trivial stationary
local equilibrium is found for
rotating massless quanta between 
two concentric cylinders of different temperatures. 
This shows that quanta may behave like a fluid 
with a B\'enard instability.

\end{abstract}

\newpage

\section{Introduction}

An understanding of nature seems to be easier 
at very small length scales
than at larger length scales.
The better the resolution power of an
observable, the less can be resolved: 
whatever the state of a physical system is, it 
cannot be distinguished from the vacuum state if the
localization regions of the observables are shrinked to a point. 
This is the content
of the Principle of Local Stability \cite{hns}, \cite{haag}.
What can be said about the state of a physical system
if the localization region of a measurement is not
completely  shrinked to a point, but is {``infinitesimally localized''}?
>From general relativity  we know that because of the
Equivalence Principle,  gravitation
is locally constant.
In \cite{hessling} a formulation 
of a Quantum Equivalence Principle (QEP)
was suggested.
According to QEP 
the states a physical system
are locally constant.

QEP was investigated in
the Rindler space-time
\footnote{The Rindler space-time is a wedge in the 
    Minkowski space-time ($ | t| < x^{(1)}$).
    It is  a simple model of a black hole.}
and it was shown
  that the Hawking--Bisognano--Wichman--Unruh
temperature \cite{hawking}, \cite{bw1}, \cite{bw2}, \cite{unruh} is
 a consequence of  QEP \cite{hessling}.

There was a discussion in the literature on
the value of the Hawking temperature of an extremal 
charged black hole.
In \cite{hawkingcoll} Hawking and collaborators
investigated the action (on the tree level) and
its topological behavior
and claimed that 
an extremal charged black hole does not have a 
definite Hawking temperature, but
equilibrium states of all temperatures may exist
outside the horizon. 
On the other hand,
Anderson and collaborators \cite{anderson} calculated the 
thermal expectation value of the energy--momentum 
tensor on the horizon and
found that 
the energy--momentum tensor is finite
only for a vanishing Hawking temperature.
The same result was derived by Moretti who
showed that only a vanishing Hawking temperature
is compatible with
QEP
\cite{moretti}.

In this article
 we continue the study of the
short-distance behavior of states in relativistic
quantum field theories. 
In section~2 it is investigated whether
the Principle of Maximum Entropy can be used
to characterize local equilibrium.\footnote{
  For different approaches to non--equilibrium see, e.g., 
  \cite{calzettahu}, \cite{umezawa}
  and the literature cited therein.}
The radiation of a hot sphere is treated
within Relativistic
Hydrodynamics in section 3.
After having collected some facts about
global equilibrium in section 4,
we formulate a 
Local Equilibrium Condition (LEC) in section 5 
and apply it
to the massless Klein--Gordon field.


\section{The Principle of Maximum Entropy}

Global equilibrium states in non--relativistic quantum systems
can be characterized by the  extremalization of
a certain functional: the entropy.
The entropy of a state $\za \cdot \ze = {\rm Tr \hat \rho (\cdot)}$
is given by
\bea
       S = - {\rm Tr} {\hat \rho} \ln {\hat \rho} 
       \label{entropy}
\eea
where $\hat \rho = \rho/{\rm Tr} \rho$ is the normalized density
matrix. Consider a system whose 
Hamilton operator $H$ is the only constant of
motion and whose total energy 
\bea
         E = \za H \ze
   \label{enerconstr}
\eea
is fixed. It is a general experience 
that, whatever the state of the
system is at some initial time,  after a long time
the system will become more and more
stationary and finally will go over into a global equilibrium
state. 
According to the  Principle of Maximum Entropy 
the system will evolve most probably into a state
of maximum entropy. 
To determine  the most probable  state 
one has to solve the following 
variational equation for the normalized density
matrix:
\bea
  \delta \left( S + c(1 - \za 1 \ze)
 + \beta(E- \za H\ze) \right) = 0,
\nonumber
\eea
where $c$ and $\beta$ are  Lagrange multipliers for 
the normalization condition ${\rm Tr}\hat \rho = 1$ and the constraint
(\ref{enerconstr}) respectively.  One finds the 
Boltzmann equilibrium distribution
\bea
        \rho = e^{-\beta H}
.\nonumber
\eea
It turns out that the Lagrange multiplier 
$\beta$ 
is just the inverse temperature of the system.

To determine the state of a system with a non--uniform 
temperature it 
was suggested \cite{mori}, \cite{zubarev0}, \cite{zubarev1}
 to apply the Principle of 
Maximum Entropy to a local form of the constraint
(\ref{enerconstr}):
\bea
          \epsilon (\vec x) = \za {\cal H}(\vec x) \ze
\label{locconstr}
\eea
where ${\cal H} (\vec x)$ is the Hamilton density 
at time zero.
This leads to  
 a continuum of Lagrange multipliers $\beta(\vec x)$ 
and 
the density matrix
\bea
     \rho_{\beta(\vec x)} = e^{-
                \int {\rm d}^3 x \beta(\vec x) {\cal H} (\vec x)}
 . \label{locden}
\eea
$1/\beta(\vec x)$ 
is interpreted as the {\em local temperature} of the
system.
For an introduction to this approach see, e.g., 
\cite{elmfors}, \cite{zubarev2}, \cite{bibi} and the
references therein.

Entropy is a probability measure for states, i.e. it assigns
a positive number to each possible state of 
the system.
The larger this number, the more
likely is a state realized. From a formal point of view it is possible
to apply the Principle of Maximum Entropy also to a system
with a local constraint (\ref{locconstr}).
But,  from a physical point of view, the
step from the global constraint  (\ref{enerconstr}) to
the local constraint (\ref{locconstr}) is 
non--trivial
and it is not clear whether 
also for local constraints  the entropy (\ref{entropy}) is the right 
probability measure to
characterize the states which are realized 
physically.
A further problem is to give a physical interpretation to
the Lagrange multipliers $ \beta(\vec x)$.

Consider the simple but important example, where the
inverse local temperature $\beta({\vec x})$  
is linear in one coordinate
\bea
  \beta(\vec x) =  
                {\beta}_0 x^{(1)}
.\label{line}
\eea
The  Hamiltonian
\bea
     H = \int {\rm d}^3 x\; x^{(1)} {\cal H} (\vec x)
\label{htild}
\eea
is the generator of a boost transformation in 
$x^{(1)}$--direction, which transforms 
observables along orbits
of constant acceleration in the Rindler spacetime.
In \cite{hns}, \cite{hessling} the short distance behavior
of the state given by Eq.~(\ref{locden}) und  Eq.~(\ref{line}) 
was analyzed for a Klein--Gordon field 
and it was found that the only temperature which is 
allowed physically, is
$\beta_0 = 2\pi$.

This result shows that
the Principle of Maximum Entropy 
is not 
strong enough to exclude the unphysical temperatures
$1/\beta_0 \neq 1/2\pi$.
The root of the problem is
that the state (\ref{locden})
is only the solution to a formal
problem, namely to find a state which extremizes 
the entropy functional (\ref{entropy})
in the case of a local constraint (\ref{locconstr}).

In summary, the Principle of Maximum
Entropy despite its success in 
characterizing global equilibrium seems, in general not to be 
a good starting point to characterize 
local equilibrium.


\section{Relativistic Hydrodynamics}

Elements of
relativistic hydrodynamics are briefly reviewed, see e. g. 
\cite{neugebauer}, and
an ultrarelativistic perfect fluid 
outside a hot sphere is analyzed.

\subsection{Thermodynamical Laws}

The stress--energy tensor of a perfect fluid reads
\bea
       T^{\mu\nu}  = (\rho + P) u^\mu u^\nu - P \eta^{\mu\nu}
\nonumber
\eea
where $u$ is the normalized velocity 4--vector of the fluid.
In the case of a radiation fluid the
equation of state $P=P(\rho)$ between its pressure
$P$ and its density $\rho$ is given by
\bea
        P = \rho / 3.
\nonumber
\eea
The 
stress--energy tensor is used to formulate
the thermodynamical laws.

\begin{description}
\item[First Law of Thermodynamics:] 

        The
        stress--energy tensor
        is  conserved
          $$
                   \partial_\nu  T^{\mu\nu} (x)= 0.
          \label{relstressconserv}
          $$
\end{description}

\begin{description}
\item[Second Law of Thermodynamics:] 
     Let 
     $
           S^\mu(x) \equiv T^{\mu\nu}(x) u_\nu(x)/T(x)
     $
     be the entropy current four-vector with respect to the
     temperature $T(x)$.
     The entropy production 
     \bea
               \sigma(x) &  \equiv  &  \partial_\mu {S^\mu}(x) \ge 0 
     \nonumber
     \eea
     cannot be negative.
\end{description}

\noindent
The conservation equations of the stress--energy tensor are also known
as the
(relativistic) Navier--Stokes equations.

The minimum requirement for a fluid 
to be in a  {\em local equilibrium} 
is a vanishing entropy production.

For a  radiation fluid in a local
equilibrium we obtain by taking the Navier-Stokes equations into account 
\bea
       \partial_\mu \ln \left( \frac{P}{T^4} \right) u^\mu = 0.
\label{mom_temp}
\eea

\subsection{Hot Sphere}

Consider a hot sphere of radius $r_0$
with the surface temperature $T_0$ which is
immersed in a radiation fluid of temperature $T_\infty$.
Can this system be in a stationary local equilibrium?

Before we consider this question within Relativistic 
Hydrodynamics, let us determine the temperature
on a planet induced by the radiation of the sun.\footnote{
        The following derivation
        is well-known and 
        included for the convenience of the reader.}

The sun can be considered as a 
black sphere 
which radiates at a temperature
$T_0 \approx 5800 {\rm K} $.
The energy emitted per time through its surface
reads according to the Stefan-Boltzmann law
\bea
       \left( \frac{E}{t} \right)_{sun,emitted}
        = \sigma_0 {T_0}^4 (4 \pi {r_0}^2)
\nonumber
\eea
where $\sigma_0$ is the Stefan-Boltzmann constant.
A planet with radius $r_p$ at the
distance $r$ absorbs the following 
fraction of the radiation on the sunlit side 
\bea
        \left( \frac{E}{t} \right)_{planet, absorbed}
       =
          \frac{\pi {r_p}^2 }{4 \pi r^2 } 
         \left( \frac{E}{t} \right)_{sun,emitted}       
\nonumber
\eea
and reradiates to the space at the same rate
\bea 
        \left( \frac{E}{t} \right)_{planet, emitted}
       =
         \sigma_0 {T(r)}^4 (4 \pi {r_p}^2).
\nonumber
\eea
The equilibrium condition 
$ (E/t)_{planet, absorbed} = (E/t)_{planet, emitted} $
implies
\bea
        T(r) = T_0 \sqrt{ \frac{r_0}{2r} }.
\label{tempdecaywright}
\eea
For the earth this gives 
$
       T_{\rm earth} \approx 280 {\rm K}, 
$
where the following astronomical data were used: 
radius of the sun $\approx$ $7/3$ lightseconds,  
distance between the earth and the sun $\approx$ 500 lightseconds.
This simple calculation leads to a surprisingly good result,
although it neglects 
the important greenhouse effect caused by the atmosphere.

\vspace*{5mm}

 Let us introduce 
spherical coordinates
\bea
       {\rm d} s^2 =  {\rm d} t^2 
                    - {\rm d} r^2
                    - r^2 {\rm d} \vartheta^2
                    - r^2 \sin^2\vartheta {\rm d} \varphi^2
\nonumber
\eea
and make a rotationally symmetric
ansatz for the equilibrium vector
\bea
       \beta^\mu \equiv v^\mu / T
                 = ( \beta^t, \beta^r, 
                       \beta^\vartheta, \beta^\varphi)
                 = \left( A(r), B(r), 0,0 \right).
\nonumber
\eea
From Eq.\ (\ref{mom_temp})
we obtain the following connection between the pressure and the 
temperature
\bea
       P \sim T^4
\nonumber
\eea
i. e. the stress--energy tensor is of the form
\bea
     T^{\mu\nu}(x) 
      &\sim& 
          T^4(x) ( 4 u^\mu(x) u^\nu(x) - \eta^{\mu\nu} ).
\nonumber 
\eea
Contracting the Navier-Stokes equations with $\beta_\mu$ yields
\bea
     3 (T^6 \beta^\nu)_{;\nu} = T^6_{,\nu}\beta^\nu.
\label{NW_contracted}
\eea
With this result the Navier-Stokes equations read
\bea
       \frac{4}{3} T^6_{,\nu}\beta^\nu \beta^\mu
     + 4 T^6 \beta^\nu {\beta^\mu}_{;\nu}
- \eta^{\mu \nu} {T^4}_{,\nu}
= 0.
\label{stresscons}
\eea
Note that Eq. (\ref{NW_contracted}) can be 
reformulated as a conservation equation
for the entropy current 4-vector 
\bea
     (T^4 \beta^\nu)_{;\nu} = 0 
\label{law2}
\eea
showing that the entropy production
of a radiation fluid is zero.
The four Navier--Stokes 
equations~(\ref{stresscons})
are equivalent to the equation
\bea
  0  &=& (T^4 B^2 + T^2) _{,r}
\label{hot_d}
\eea
i. e.
\bea
       B^2 =  \frac{C_1}{T^4} - \frac{1}{T^2}.
\label{b2c1}
\eea
Together with the solution of Eq. (\ref{law2})
\bea
       B = \frac{C_2}{ r^2 T^4}.
\nonumber
\eea
we  obtain for
the temperature outside the sphere
\bea
        \frac{C_1}{T^4} - \frac{1}{T^2} = \frac{{C_2}^2}{ r^4 T^8}
\label{tempsphere}
\eea
The integration constant $C_1$ turns out to be
\bea
        C_1     &=& {T_\infty}^2
\nonumber
\eea
If $C_1$ is inserted into Eq. (\ref{b2c1}), 
it follows from the positivity of $B^2$
\bea
          T_\infty  \ge T(r)  
\label{Tineq}
\eea
i. e. in a local equilibrium the temperature
of the quanta at infinity cannot be smaller than the
temperature of the sphere, $T_\infty \ge T_0$.

This bound is not changed significantly 
        if gravity is taken into account by using 
        the Schwarzschild metric in the Navier-Stokes equations,
        since the right hand side of 
        Eq. (\ref{Tineq}) is simply multiplied 
        by the  factor $ [1-r_{SS}/r)]^{1/2}$
        where $r_{SS} \approx 3$ km  is the 
        Schwarzschild radius of the sun. 

To summarize, the radiation emitted by the sun cannot be 
understood as a stationary rotationally symmetric 
local equilibrium process.

\vspace*{5mm}

However, in the case of a black hole, 
$ r_0 =  r_{SS} $, 
a local equilibrium may exist 
with a temperature 
which is zero everywhere
outside the black hole. 
This follows from the fact that 
Eq. (\ref{tempsphere}) generalizes to 
\bea
        \frac{C_1}{T^4} - \frac{r-r_{SS} }{r} \frac{1}{T^2} 
       = \frac{{C_2}^2}{ r^4 T^8}.
\label{tempbh}
\eea
Since a cold black hole
is in contradiction to the Hawking effect
\cite{hawking}, we conclude
that the theory of Relatistic Hydrodynamics seems,
in general also not
to be a good starting point to
characterize local equilibrium.


\section{Global Equilibrium}

\subsection{Modular Evolution}

In a quantum mechanical system of finite degrees of freedom 
the expectation
value $\za A \ze$ of any observable $A$
in a state $\za \cdot \ze$  can be characterized by a  
density matrix $\rho$
\bea
       \za A \ze &=& \frac{
                {\rm   Tr}\, \rho A}{ {\rm Tr}\, \rho}
  .     \label{poltamann-mod}
\eea
If one introduces  the
{\em modular Hamiltonian} $\tilde H$ via
\bea 
    e^{- {  \tilde H / T}} &=& \rho
       \label{loglog}
\eea 
where the parameter $T$ is extracted for later convenience,
the {\em modular
evolution} 
\bea
  \alpha_\tau (A) = e^{i{\tilde {H}}\tau}Ae^{-i{\tilde {H}}\tau}
  \label{mod-evo}
\eea 
can be defined.

 Cyclicity of the trace gives
the  {\em KMS--condition} \cite{kubo}, \cite{ms}
\bea
        \za \alpha_\tau(A) B \ze 
         &=& \za B \alpha_{\tau + i/T }(A) \ze
         \label{kms-mod}.
\eea

In quantum field theory the right hand side of 
  Eq.~(\ref{poltamann-mod})
does not
exist, but the 
KMS--condition  (\ref{kms-mod})
can be used directly to characterize the state 
\cite{hhw}, \cite{haag}.\footnote{
  Strictly speaking, the modular Hamiltonian $\tilde H$ cannot be
  defined as the logarithm of the density matrix, as in
   Eq.~(\ref{loglog}), but the necessary 
  technical modifications to make
  $\tilde H$ well--defined (see e.\ g.\ \cite{haag}) 
  are not important for our
  purpurses.}

By using the Fourier transformation of the KMS condition
(\ref{kms-mod}),
 it is possible to represent a 
state 
in terms of a commutator
\bea
 \langle B \alpha_{i \varepsilon}(A) \rangle 
 & = &
         \frac{-1}{2\pi}
         \int {\rm d} \omega
         {\rm d} \tau
         \langle [B,\alpha_{\tau}(A)] \rangle 
         \frac{  e^{-i\omega (\tau- i \varepsilon)}}
         { e^{- \omega/T}-1}    
                                       \nonumber \\
 & = &
         \frac{T}{2i}
         \int {\rm d} \tau
         \langle [B,\alpha_{\tau}(A)] \rangle 
         \coth (\pi T(\tau- i \varepsilon))   .   
\label{kms-simple}
\eea
Here, it is assumed that $ \langle B \alpha_\tau(A)  \rangle$
goes to zero in the limit $|\tau| \rightarrow \infty.$\footnote{
   Otherwise, in Eq.~(\ref{kms-simple}), one can replace
   the left hand side  by
   $ \langle B \alpha_\tau(A)  \rangle - 
    \langle B\rangle \langle A \rangle
   $; see e.g. \cite{hns}, \cite{haag}.} 
$\varepsilon$ is a tiny positive number which is 
necessary to give  Eq.~(\ref{kms-simple})
a
well--defined meaning in the sense of 
distributions and which finally goes to zero. 
The essential point is that
the right hand side of  Eq.~(\ref{kms-simple})
can be used
to calculate states in linear field theories
for a given modular Hamiltionian, since
the commutator of bosonic fields and
the anti--commutator of fermionic fields are 
multiples of the unit operator and thus independent of the state .

Global equilibrium states  $\za \cdot \ze_{\rm eq}$ are
characterizable by the 
modular Hamiltonian 
\bea 
  \tilde H =  u_\mu P^\mu
\label{modHu}
\eea
where $P^\mu$ are the time and spatial translation generators and
$u$ is the {\em equilibrium velocity} 
\bea
                 u^\mu u_\mu = 1.
\nonumber
\eea
The parameter $T$ introduced in  Eq.~(\ref{loglog}),
can be identified with the
temperature of the equilibrium state. 
It is convenient to introduce the
{\em equilibrium vector} 
\bea
  \beta^\mu = u^\mu / T . 
\nonumber
\eea

Often a co--moving coordinate system is chosen 
where the equilibrium velocity is at rest
\bea
                u^\mu = (1,0,0,0)
\nonumber
\eea 
called {\em rest system} of the heat bath.
With respect to this system the ``fluid of quanta'' 
appears to be stationary and spatially isotropic
and the modular Hamiltonian reduces to the
Hamilton operator.


\subsection{Massless Klein--Gordon Field}

The massless Klein-Gordon field $\phi(x)$
in Minkowski spacetime 
is a solution of the wave equation
\bea 
             \eta^{\mu\nu} \partial_\mu \partial_\nu \phi(x) 
   =
            \phi_{,tt} -
            \phi_{,x^{(1)}x^{(1)}} -
            \phi_{,x^{(2)}x^{(2)}} -
            \phi_{,x^{(3)}x^{(3)}} 
            = 0.
\nonumber
\eea
The commutation relation reads
\bea
   [\phi(x),\phi(x')] = -i D_u(x-x') .
\label{ccr}
\eea 
$D_u(x)$ is the massless Pauli--Jordan commutator function
\bea
   D_u(x-x') = \frac{1}{4\pi } {\rm sign}(ux-ux')
    \delta( \sigma (x,x') )
\nonumber
\eea 
where $u$ is a timelike vector and
\bea
          2  \sigma(x,x') = 
                           (t-t')^2 - (\vec x - {\vec x}')^2
\nonumber
\eea
is the square of the geodesic distance between the points
$x$ and $x'$.

\subsubsection{Thermal 2--Point Function}
In this subsection we use 
 Eq.~(\ref{kms-simple}) 
to calculate the 2--point function
of a massless Klein--Gordon field $\phi(x)$ 
in a global equilibrium state.

The modular Hamiltonian (\ref{modHu}) 
    generates 
a timelike evolution in the direction of the 
equilibrium volocity $u$
\bea
         \alpha_\tau \phi(x) \equiv \phi_\tau (x) =  \phi(x+\tau u) 
 .   \label{timetrans}
\eea

To
 calculate the 2--point function $\za \phi(x') \phi(x) \ze_{\rm eq}$
of a global equilibrium state $\langle \cdot \rangle_{\rm eq}$,
we make use of Eq.~(\ref{kms-simple})
and the commutation relation (\ref{ccr})
and obtain
\bea
  \za \phi(x') \phi_{i \varepsilon}(x) \ze_{\rm eq} =
  \frac{T}{8 \pi  } \left(
 \frac{1}{{\dot \sigma}_+}
 \coth (\pi T (\tau_+ - i\varepsilon))
 -
 \frac{1}{{\dot \sigma}_-}
 \coth (\pi T (\tau_- - i\varepsilon))
 \right)
 \label{result}.
\eea
The times $\tau_\pm$ are implicitly given as the solutions of the
equation
\bea
      2 \sigma_\tau(x,x') \equiv  (\tau u  - \Delta x)^2 = 0
\label{geod}
\eea
where
\bea
                \Delta x = x' - x.
\nonumber
\eea
$\tau_\pm$  
mark the two times
at which   the orbit of 
the one--parametric diffeomorphism
\bea
           x \rightarrow x + \tau u
\nonumber
\eea
becomes lightlike to the point  $x'$. 
${\dot \sigma}_\pm$ are the absolut values of the
time derivative of  Eq.~(\ref{geod})
at the times $\tau_\pm$:
\bea
     {\dot \sigma}_\pm = \left| \frac{{\rm d}}{{\rm d}\tau}
      \sigma_\tau(x,x') \right|_{\tau = \tau_\pm}
\label{ablgeod}.
\eea
We find
\bea
    \tau_\pm = {u {\cdot} \Delta x \pm {\dot \sigma}}
\label{tau1s}
\eea
and
\bea
   {\dot \sigma_+} =  {\dot \sigma_-} 
  =  \sqrt{ ( u {\cdot} \Delta x)^2 -  (\Delta x)^2 } 
 \equiv {\dot \sigma}
.\label{sig1s}
\eea
${\dot \sigma}$ is real,
because the equilibrium vector is timelike.

In summary, the thermal 2--point function of 
a massless Klein-Gordon field in a global equilibrium state
is characterized 
by  Eqs.~(\ref{result}), (\ref{tau1s}), (\ref{sig1s}).

\subsubsection{Energy--Momentum Tensor}

The canonical energy--momentum tensor of a massless Klein--Gordon
field in the Minkowski spacetime is formally given by
\bea
             T_{\mu \nu} (x)
          &=&
             \partial_\mu  \phi (x) \partial_\nu  \phi (x)
             -
             \frac{1}{2}\; \eta_{\mu \nu} 
             \partial^\rho \phi (x) \partial_\rho \phi (x).
\eea
It can be rewritten as
\bea
                 T_{\mu \nu} (x)  
          &=&
             - \phi(x) \partial_\mu \partial_\nu \phi (x)
             + \left(
             \frac{1}{2}\partial_\mu \partial_\nu 
             -         
             \frac{1}{4} \eta_{\mu \nu} 
             \partial^\rho \partial_\rho \right) \phi^2(x).
\label{Tformal}
\eea

Because of the distributional character of quantum fields, 
the product of two fields at the same spacetime point 
has to be defined by a regularization prescription.
According to the {\em Principle of Local Definiteness}
physical states have the same ``singularity structure''
at short distances
(see \cite{haag} for details) and, thus, 
different physical regularization prescriptions
differ only in their finite parts.

We define the energy--monentum tensor of a 
state $\za \cdot \ze$ relative to the 
vacuum state $\za 0| \cdot |0\ze$
\bea
             T_{\mu \nu} (x)
             &=&
               D_{\mu \nu}
               \left( 
                      \frac{}{}
                      \za \phi(x') \phi(x) \ze 
                    -
                      \za 0| \phi(x') \phi(x) |0 \ze 
               \right) 
\label{TMink}
\eea
where we introduced the differential operator
\bea
             D_{\mu \nu}
             = 
              - \lim_{x' \rightarrow x} \partial_\mu \partial_\nu
              + 
                (\frac{1}{2}\partial_\mu \partial_\nu 
                 -         
                 \frac{1}{4} \eta_{\mu \nu} 
                 \partial^\rho \partial_\rho  
                )   
                 \lim_{x' \rightarrow x}
.
\label{Dmunu}
\eea

The differential operators $\partial_\mu\partial_\nu$
in $D_{\mu \nu}$ act on the point $x$,
i.~e.\ on only one point of the 2--point function, contrary
to the standard point--splitting method which is based 
on non--local differential operators \cite{dwb}, 
\cite{christensen}. 
The details of the regularization (\ref{TMink}), 
in particular the interplay between limiting processes und
derivatives, originate in the second line
of Eq.~(\ref{Tformal}). Note that the energy-momentum tensor
is symmetric.

The  2--point function of the vacuum state 
is roughly inverse proportional to
the square of the geodesic distance, more precisely it reads
\bea
   \za 0| \phi(x') \phi_{i \varepsilon}(x)|0 \ze 
 &=&
   \frac{1}{2 \pi^2 } \frac{-1}{\sigma_{i\varepsilon}(x',x)}
\label{vacc}
\eea
where
\bea
   2 \sigma_{i\varepsilon}(x',x) =
  (x - x' + i \varepsilon u)^2 .
\label{sigmaieps}
\eea
is the square of the 
geodesic distance between the points $x$ and $x'$ modified
by an $i\varepsilon$--prescription, where $u$ is a timelike 
vector
which we identify with the equilibrium
velocity.

To calculate the energy--momentum tensor, we 
expand the 2--point function of the equilibrium state
(\ref{result}) around $x' \approx  x $, i.\ e.\
$\Delta x \approx 0$ or, equivalently, $\tau_\pm \approx 0$.
By using
\bea
               \coth \varepsilon =  \frac{1}{\varepsilon} 
                        + \frac{\varepsilon}{3} 
                        + \frac{\varepsilon^3}{45} 
                        + O(\varepsilon^5)
\nonumber
\eea
we find
\bea
   \za \phi(x') \phi_{i \varepsilon}(x) \ze_{\rm eq} 
 &=&
   \frac{1}{4 \pi^2 } \frac{-1}{(\Delta x - i \varepsilon u)^2} 
 +
   \frac{1}{12 \beta^2} 
 \nonumber \\
 & & 
 -
   \frac{\pi^2}{180}
   \left( T^6 4 \beta_\mu \beta_\nu - \eta_{\mu\nu} T^4 \right) 
    (\Delta x)^\mu(\Delta x)^\nu
 +
   O( \Delta x)^3 
\label{resultexpand}
\eea
where 
$
             \beta^\mu = {u^\mu}/{T}
$
is the {\rm equilibrium vector}.
Inserting in Eq.~(\ref{TMink}) the 
Eqs. (\ref{resultexpand}) and (\ref{vacc})
we obtain
\bea
             {T^{\rm (eq)}}_{\mu \nu}
             =
             \frac{ \pi^2 }{90} 
             \left( T^6 4 \beta_\mu \beta_\nu - \eta_{\mu\nu}T^4 
             \right)
\label{Tvac}
\eea
The energy momentum tensor of a global equilibrium
is conserved
          \bea  
                \partial_\nu  {T^{\rm (eq)}}^{\mu    \nu} (x) = 0
          \label{Tconserv}
          \eea
simply because the components of ${T^{\rm (eq)}}_{\mu \nu}$ 
are constant, and traceless
          \bea
                             {{T^{\rm (eq)}}_{\mu}}^{\mu} (x) = 0 . 
          \label{Ttrace}
          \eea

In the rest system of the heat bath
\bea
          \beta^{\tilde \mu}  = \frac{1}{T} (1,0,0,0)
\nonumber
\eea
the energy--momentum tensor simplifies to
\bea
             {T^{\rm (eq)}}_{\tilde \mu \tilde \nu}
             =
             \frac{ \pi^2 }{90} T^4 {\rm diag}(3,1,1,1)
\nonumber
\eea 
i.~e. the energy density is proportional
to the fourth power of the temperature
\bea
                {T^{\rm (eq)}}_{\tilde 0 \tilde 0}
              =\frac{\pi^2}{30}
                 T^4
\nonumber
\eea
which is the Stefan--Boltzmann law.
In other words,
Eq.~(\ref{Tvac}) is a ``heat bath frame--independent''
formulation of the Stefan--Boltzmann law.


\section{Local Equilibrium}

Global equilibrium is well understood. Descriptions
exist at any level of rigor. Local equilibrium, 
on the other hand,
is much less studied despite the fact that
it is much more often realized in nature
than global equilibrium.

In this section we present a formulation of 
local equilibrium  as a state
which  cannot be distinguished from
a global equilibrium 
state by ``infinitesimally localized measurements''.


\subsection{Short Distance Behavior of States}

If the localization region of an  observable $A$ is made smaller
 and
smaller,
the expectation value $\za A \ze$ of an observable $A$ in a 
local equilibrium state $\za \cdot \ze$
should become 
more and more identical to the expectation value 
$\za A \ze_{\rm eq}$
of 
$A$ in a global equilibrium state $\za \cdot \ze_{\rm eq}$.
We assume that the global equilibrium is characterizable
by the temperature $T$ and the equilibrium 
velocity $u$.

The shrinking of the localization region of an observable
can be described by a one--parametric scaling 
procedure $\delta_\lambda$ \cite{fh}, \cite{haag}. 
Let $\phi(x)$ be a Klein-Gordon field. The 
one--parametric map $\delta_\lambda$ scales
the amplitude of the field $\phi(x)$ by a factor
$N_\phi (\lambda)$ and shifts its 
localization point along a path $\eta_\lambda x $
\bea
      \delta_\lambda \phi(x) 
     = 
      N_\phi (\lambda) \phi( \eta_\lambda x ).  
\nonumber
\eea

The path $\eta_\lambda x $ is defined as the 
1--parametric diffeomorphism
\bea
    (\eta_\lambda x)^\mu ={ x_*}^\mu + \lambda
                                   ( x^\mu - {x_*}^\mu )
     \label{lammel}
\eea
which has  the properties $\eta_1 x = x$
and $\eta_0 x = x_*$.
In the limit $\lambda \rightarrow 0$ the
localization point $x$ is
scaled into the point
$x_*$ along a straight line.

The {\em  scaling function} $N_\phi(\lambda)$
is determined relative
to a state $\za \cdot \ze$. 
It is positive and monotone and  
has to be adjusted in such a way
that the 
{\em scaling limit} 
\bea
  \lim_{\lambda \rightarrow 0} \za \delta_\lambda 
   (\psi_1(x_1) \dots \psi_n(x_n)  \ze
\nonumber
\eea
exists for all $n$--point functions
 and is non--vanishing for some. 
The linear scaling function $N(\lambda)= \lambda$ is
a suitable scaling function 
for the Klein--Gordon field $\phi(x)$
in a thermal state, as can directly be seen 
by studying the short distance behavior of
Eqs.\ (\ref{geod}).
Each derivative increases the scaling function by one
power in the scaling parameter, 
e.~g.\ $N_{\partial_\mu \phi}(\lambda)= \lambda^2$.

In non--linear quantum field theories the 
scaling function
can be
determined by renormalization group techniques, 
e.\ g.\ \cite{hessling}.

An important statement about the
scaling limit is made by the 
{\em Principle of Local Stability} (PLS) \cite{hns}, \cite{haag}.
The fact, that the scaling function of a thermal
Klein--Gordon field does not depend 
on the temperature, is not an  accident
but a general property of physical states.
\begin{description}
\item[PLS:]
  The 
  scaling limit of any given physical state 
  $\za \cdot \ze $ agrees with the
  scaling limit of the vacuum state 
  $ \za 0| \cdot | 0  \ze$ in the Minkowski spacetime
  \bea
     \lim_{\lambda \rightarrow 0} \za \delta_\lambda A  \ze
     =
     \lim_{\lambda \rightarrow 0} \za 0| \delta_\lambda A |0 \ze
  \nonumber
  \eea
  for all observables $A$.
\end{description}

According to the {\em Quantum Equivalence Principle} (QEP) 
\cite{hessling} the scaling limit does not change locally.
\begin{description}
\item[QEP:]
  For physical states $\za \cdot \ze$ the expectation value of 
  any scaled observable
  $\za \delta_\lambda A \ze$
  is locally  constant  
  around the scaling point~$x_*$. \\
  For linear theories 
  and with respect to
  an inertial coordinate system\footnote{
               If non--inertial coordinates are used 
               around the scaling point,  
              the derivative ${\rm d}/ {\rm d} \lambda$
              in Eq.\ (\ref{qap2}) has to be replaced by a 
              covariant 
              derivative, see \cite{hessling}.
  }
  this condition 
  can be written as the 
  extremum condition
  \bea
    \lim_{\lambda \rightarrow 0}
                                  \frac{d}{d \lambda}
                 \za \delta_\lambda A
      \ze 
          &=& 0
      \label{qap2}.
  \eea
  The scaling limit, 
  $ 
   \lim_{\lambda \rightarrow 0} \za \delta_\lambda A \ze,
  $
  is a continuous function in the scaling point~$x_*$. 
\end{description}

In other words, 
the first non--trivial information about the state of 
a linear quantum field  is beyond the first
order in the scaling parameter $\lambda$. 
In the next subsection we investigate whether
properties of local equilibrium
become visible at the second order 
in the scaling parameter.

\subsection{Local Equilibrium Condition}

In studies of the    
local properties of a quantum statistical system
people often consider the limit
of ``small, but not too small volumes''.
Intuitively it seems to be clear that,
if a volume is sufficiently small, the
physical process of interest appears to 
be homogenous. On the other hand, if there
are not enough particles included in a volume  
the   ``statistical average'' is not
well--defined.
The existence of such volumes is expected 
phenomenologically.
Analogously,
in the {\em coarse graining method} the
degrees of freedom of a system are integrated out up to
a ``characteristic length scale'';
the remaining degrees of freedom
are treated as phenomenological parameters.

Although the phrases ``small, but 
not too small'' or ``characteristic length scale''
are quite intuitive, a general strategy to characterize
 them conceptually well founded
seems not to be known.

It is  our point of  view 
to  look as closely as possible to the 
details of a system. 
Therefore, we propose to go over to the limit
of infinitesimally small volumes or 
characteristic length scales,
respectively. In the {\em Local Equilibrium Condition} (LEC) 
this idea is formulated quantitatively.

\begin{description}
\item[LEC (Part 1):]
  A state $\za \cdot \ze$  is in a local equilibrium
  in a given point $x_*$ of a physical system,
  if  it can be approximated by  
  a global equilibrium state $\za \cdot \ze_{\rm eq}$ of 
  a certain temperature
  $T_*=T(x_*)$ and a certain equilibrium velocity $u_*=u(x_*)$
  up to the second order in the scaling parameter $\lambda$
  \bea
    \lim_{\lambda \rightarrow 0}
                                  \frac{d^2}{d \lambda^2}
    \left( 
       \frac{}{}
                 \za \delta_\lambda  A  \ze 
      -           \za \delta_\lambda A  \ze_{\rm eq} 
    \right)
          &=& 0
  .    \label{lec}
  \eea
\item[LEC (Part 2):]
  The energy-momentum tensor of the state $\za \cdot \ze$
  defined in Eq. (\ref{TMink}),
  is identical to the energy-momentum 
  tensor of the global equilibrium state
  \bea
          T_{\mu\nu}(x_*) = {T^{\rm (eq)}}_{\mu\nu}(x_*)
     \nonumber    
  \eea
  in such a way that it fulfills the Navier-Stokes equations
  \bea
         \partial^\nu T_{\mu\nu}(x_*) = 0 
  \label{navier-stokes}
  \eea 
  and the entropy production is vanishing 
  \bea        
          \sigma( x_*) \equiv \partial_\mu S^\mu (x_*) = 0
  \nonumber
  \eea
  where $S_\mu = T_{\mu\nu} \beta^\nu$ 
  is the entropy current 4-vector.
\end{description}

The parameters of the local equilibrium, i. e. 
the local temperature $T_*$
and the local equilibrium velocity $u_*$,
are functions of the scaling point $x_*$.
These functions are also constrained by
the field equations, as can be seen below.

To test whether LEC is sensible to characterize local equilibrium, 
we apply it to an important class of states, the Hadamard states.


\subsubsection{Hadamard States of Massless Scalar Quanta}
 
Hadamard states of a linear scalar field
are quasifree states\footnote{
            A state is called quasifree, if its truncated
            $n$--point functions vanish for $n\neq 2$.}
with a  specific singularity structure:
the   2--point function is identical
with Hadamard's fundamental solution of the wave equation
\cite{dwb}
   \bea
     \za  \phi(x')  \phi_{i \varepsilon} (x)  \ze
     &=& \frac{-1}{8 \pi^2} \left(
     \frac{U}{\sigma_{i\varepsilon}} + V \ln \sigma_{i\varepsilon}
        \right)
        + W 
   \label{hadamard}
   \eea
where $2 \sigma_{i\varepsilon}$ is the square of the
geodesic distance between the points $x'$ and $x$, 
as defined in  Eq.~(\ref{sigmaieps}).
$U, V, W$ are regular functions in $x$ and $x'$.
The information  about  the state is contained in $W$; $U$
and $V$ are state--independent and are uniquely fixed by 
the geometry of the space-time.
In a flat spacetime they read \cite{dwb}
\bea
               U &=& 1  \nonumber \\
               V &=& - \frac{1}{2} m^2 + \frac{1}{8} m^2 \sigma
                      + O(\sigma^{3/2}) 
\label{uv}
\eea
where $m$ is the mass of the Klein--Gordon field
which is assumed to be zero.

An application of the first part of LEC to Hadamard states yields
\bea
          W(x_*,x_*) = \frac{1}{12} {T_*}^2.
  \label{wwbe}
\eea
To see this, make use of  Eq.~(\ref{resultexpand})
and take into account that
the state--independent singular parts,
 $1/\sigma$ and $\ln \sigma$, cancel
because of the difference in
(\ref{lec}), 
and that the scaled state--dependent part
$W(\eta_\lambda x',\eta_\lambda x)$ 
is regular in the limit $\lambda \rightarrow 0$.
This means
that the derivative condition of LEC does not depend on 
the scaling function  $\eta_\lambda$ in the sense 
that  Eq.~(\ref{lammel}) can be replaced by any 
one--parametric scaling diffeomorphism $\eta_\lambda$ 
which has $x_*$ as a fix--point.
In other words, LEC is independent on the
choice of the coordinate system around the
the scaling point $x_*$. 

Hadamard states with a state--dependent function
which is positive on the diagonal $x=x'=x_*$
\bea
       W(x_*,x_*) \ge 0
\nonumber
\eea
have the {\em local temperature}
\bea
       T_* = \sqrt{ 12 \, W(x_*,x_*)}
\label{loctemp}
\eea
in the scaling point $x$.

To study the implications of the field equations,
we make the following ansatz for the 2--point function
of a local equilibrium state
\bea
     \za  \phi(x')  \phi(x)  \ze_{\rm leq}
     &=&
        \frac{-1}{8\pi^2}  
        \frac{1}{\sigma_{i\varepsilon}}
     +
        \frac{1}{12} T^2(x)
        \nonumber \\
 & &    
     +
        \; W_{\mu}(x) \Delta x^\mu
     + 
        \frac{1}{2} \; W_{\mu \nu}(x) \Delta x^\mu \Delta x^\nu
     +
        O(\Delta x^3) 
\label{locexpand2}
\eea
where we have already taken into account the 
result~(\ref{loctemp}).
The equation of motion for 
the field $\phi(x')$ yields
\bea
            \partial_{\mu'} \partial^{\mu'}
           \za \phi(x') \phi(x) \ze
           = 0
\nonumber
\eea
and it follows
\bea
            {W^\mu}_\mu (x) = 0
\label{LECconstraint1}
\eea
i.~e.\ the coefficients $W_{\mu\nu}$ are traceless.
From the field equation for $\phi(x)$ we obtain
an inhomogenous wave equation 
\bea
           \partial_{\mu} \partial^{\mu} T^2(x)
            = 24 {W^\mu}_{,\mu}
\label{LECconstraint2}
\eea
for the square of the 
{\em local equilibrium temperature}  $T(x)$.

The energy-momentum
tensor takes the form
\bea
        T_{\mu\nu} (x)
     &=&
         \frac{\pi^2}{90} 
         \left(
                T^6(x) 4 {\beta}_\mu(x)  {\beta}_\nu(x)
- \eta_{\mu\nu} T^4(x) 
        \right)
\label{tmunubetw}
\eea
if the local equilibrium exists not only in a point
but in a region of the spacetime. This
follows from the second part of LEC and 
energy-momentum tensor Eq. (\ref{Tvac}) of 
the global equilibrium state.
On the other hand we obtain 
\bea
      T_{\mu\nu} (x) 
    =
      - \frac{1}{2} W_{\mu\nu} (x)
+ W_{\mu,\nu}(x)    + W_{\nu,\mu}(x)
- \frac{1}{12} {T^2(x)}_{,\mu\nu}
- \frac{1}{48} \eta_{\mu\nu}  
  {{T^2(x)}_{,\rho}}^{,\rho}
\nonumber
\eea
by combining Eq. (\ref{TMink}) and Eq. (\ref{locexpand2}). 
The trace 
\bea
     {T_\mu}^\mu (x)
    =
- \frac{1}{12} {T^2(x)_{,\rho}}^{,\rho}
\nonumber
\eea
has to vanish since the right hand side
of Eq. (\ref{tmunubetw}) has this property, 
i. e. we arrive at a homogenous wave equation
\bea
       \partial^\mu \partial_\mu T^2(x) = 0 
\label{tempwave}
\eea
for the square of the local equilibrium temperature.
Consequently the coefficients $W_\mu(x)$ fulfil
\bea
          \partial_\mu W^\mu(x) = 0.
\label{LECconstraint3}
\eea

Note that LEC does not lead to further constraints
if it is applied to derivatives of fields.\footnote{
     This remark is of importance if LEC is applied to
     free photons since a basic photon observable, 
     the field-stress tensor
     $F_{\mu\nu}$, is built from derivatives of the
     unphysical vector potential $A_\mu$
     whose thermal 2-point functions have the
     form of Hadamard states. However, it is 
     possible to construct a photon observable
     in such a way that LEC is sensitive to it.
     LEC in free gauge theories 
     will be considered
     in a separate publication.
    }
To see this, note that
the 2-point function
$\za \phi(x') \partial_\mu \phi(x) \ze $
has the scaling function $\lambda^3$, i. e. 
the derivative condition in first part of LEC 
is not sensitive to it.

Now, we are in a position to clarify the meaning
of an ``{\em infinitesimally localized measurement}'': 
it is the measurement of any scaled observable in the vincinity 
of a given spacetime point up to the
second order in the scaling parameter~$\lambda$
and a determination of the energy-momentum tensor
in that point.

\vspace*{10mm}
In summary, LEC leads to a set of constraints 
for the short distance behavior of states.
The local equilibrium vector $\beta^\mu(x)$
has to fulfil the Navier-Stokes equations and 
the square of the local temperature 
$T^2(x) = 1/\beta^\mu(x) \beta_\mu(x)$
has to fulfil a wave equation, i. e. we obtain 
an overdetermined system of five partial differential equations
for the four components of the local equilibrium vector.
The wave equation (\ref{tempwave}) describes quantum effects 
not visible within Relativistic Hydrodynamics.

\vspace*{5mm}

In \cite{hessling} it was shown that the derivative condition
(\ref{qap2}) has to be modified if one wants to 
formulate a Quantum Equivalence Principle
for asymptotically free quantum field theories, since
in QCD the running coupling constant does not
smoothly go to zero in the short distance limit $\lambda \rightarrow
0$, but $\sim 1/\ln \lambda$. 
Because of the same reason
 the derivative condition in the first part of LEC, 
(\ref{lec}) needs
a modification if one wants to characterize local
equilibrium states 
for self--interacting quantum fields.
As already mentioned in \cite{hessling}, 
the necessary modifications might be 
found by calculating  the short distance
expansions of $n$--point functions via
renormalization group techniques.
In this context the ``wave front sets'' \cite{ratzi1}
could be of importance (see also
\cite{ratzi2}, \cite{brunetti}, \cite{fredue} and the 
literature cited therein).

\subsubsection{Killing Fields and the Unruh--Effect}

The Navier-Stokes equations are solved if
the local equilibrium vector $\beta^\mu(x)$
is a Killing field, i.~e. if it is a solution
of the equation
\bea
         \beta_{\mu,\nu}+\beta_{\nu,\mu}=0.
\nonumber
\eea
This can most easily be seen if Eq.~(\ref{stresscons})
is rewritten as
\bea
    ( \beta_{\mu,\nu} +\beta_{\nu,\mu}) \beta^\nu
   + 
    \beta_\mu {\beta^\nu}_{,\nu}
   = 
     3 \beta_\mu  
        ( \beta_{\rho,\nu} +\beta_{\nu,\rho} ) \beta^\nu \beta^\rho
        / \beta^2 .
\label{convkill}
\eea

The simplest timelike Killing vector is
\bea
               \beta^\mu = \frac{1}{T_0} (1,0,0,0)
\nonumber
\eea
where $T_0$ is a constant.
It characterizes a global equilibrium with
the temperature $T_0$
in the Minkowski spacetime.

Not all Killing vectors are compatible to LEC and, thus, 
do correspond to
a local equilibrium. The non-stationary  Killing vector
\bea
               \beta^\mu = \frac{1}{T} (x^{(1)},t,0,0)
\nonumber
\eea
is timelike in the Rindler  wedge ($|t|<x^{(1)}$).
It 
fulfills the wave equation condition~(\ref{tempwave}) only if $T=0$.
The temperature $T=0$  
is not only 
the temperature of the Minkowskian 
vacuum state, but of special importance.
It is exactly the
Bisognano--Wichmann--Unruh temperature, see e.~g. \cite{haag},
although not written in its standard form but expressed
with respect to an inertial time coordinate.
We derived it by applying LEC to an arbitrary point
{\em inside} the Rindler wedge. Thus, 
LEC is sensitive to physical structures which cannot
be resolved by PLS or QEP.

In phenomenological fluid mechanics,
a system with a vanishing entropy production
is considerd to be in an 
``(incomplete) equilibrium'', see e.~g. \cite{neugebauer}.
According to this point of view, 
each local equilibrium vector $\beta^\mu(x)$ 
which is a Killing vector,  
would correspond to an ``(incomplete) equilibrium''. 
Our derivation of the Bisognano--Wichmann--Unruh
temperature shows that this point of view is not
justified in 
general. 


\subsubsection{Hot Sphere}

The radiation from a hot sphere was already considered in Sec. 3
within Relativistic Hydrodynamics. 

We now show that, as a consequence of LEC,
massless scalar quanta outside a hot sphere 
can only be in a  
stationary radially symmetric local equilibrium,
if they are in a global equilibrium.

Because of the rotational symmetry  we introduce 
a spherical coordinate system and assume that  
the temperature $T = T( r)$
is a function of only  the radial coordinate. 
The statement follows from the fact that
the wave
equation (\ref{tempwave}) gives 
for the square of the local temperature 
\bea
       T^2 (r ) = \frac{a_0}{r} + a_1   
\label{tempdecay}
\eea 
where $a_0$ and $a_1$ are integration constants, 
which contradicts the prediction (\ref{tempsphere})
of the Navier-Stokes equations, in general.\footnote{
     It is amazing to note that in the Eqs. 
      (\ref{tempdecay}) and  (\ref{tempdecaywright})
     the local temperature shows the same dependence 
     on the distance,  $T(r) \sim r^{-1/2} $.  
     }


\subsubsection{Hot Cylinder}

Consider massless scalar quanta between two
concentric cylinder with different surface temperatures.

It is convenient to introduce axially symmetric coordinates
\bea
       {\rm d} s^2 =  {\rm d} t^2 
                    - {\rm d} r^2
                    - {\rm d} z^2
                    - r^2 {\rm d} \varphi^2.
\nonumber
\eea
For the equilibrium vector we make the ansatz
\bea
       \beta^\mu \equiv v^\mu / T
                 = ( \beta^t, \beta^r, 
                       \beta^z, \beta^\varphi)
                 = \left( A(r), 0, 0, B(r )  \right).
\nonumber
\eea
The Navier-Stokes equations reduce to one equation   
\bea
            B^2 = \frac{T^2_{,r}}{2 r T^4}.
\label{Bcylinder}
\eea
The wave equation (\ref{tempwave}) is solved by
\bea
        T^2( r) =  {T_0}^2 
                   - ( {T_0}^2 - {T_1}^2 )
                  \frac{\ln r/r_0}{\ln r_1/r_0}    
\nonumber
\eea
where we assumed that the cylinder of radius $r_0$
has the temperature $T_0$ and the cylinder of radius $r_1 > r_0$
the temperature $T_1$.
If this result is inserted into (\ref{Bcylinder}) it follows from
the condition that the components of the local equilibrium vector
have to be real,
\bea
         T_0 \le T_1
\nonumber
\eea
i. e. the temperature of the outer cylinder cannot be smaller
than the temperature of the inner cylinder in the case of 
a local equilibrium.

In summary, we have found a non-trivial stationary local equilibrium
vector which characterizes rotating quanta around the
symmetry axis
since $B=\beta^\varphi$ is non-vanishing.

With respect to this 
``structure generating behavior'' 
the gas of scalar quanta  seems to be 
comparable 
to  a horizontal layer of fluid
which is heated from the bottom. If the temperature
of the fluid on the top surface
is fixed  and if the temperature gradient
is stronger than a critical 
value, a {\rm B\'enard instability}
is observed: the fluid starts to rotate in 
macroscopically regular patterns  and, for
example,  horizontal
roller can be observed \cite{haken}.



\vspace*{1cm}

{\em 
This work is an extended version of \cite{hesslinglec}
where it was suggested to characterize 
local equilibrium as a state which cannot be
distinguished from a global equilibrium state
within a sufficiently small spacetime region. 

After we had derived the material presented here,
Buchholz, Ojima and Roos published 
a general scheme to characterize local
equilibrium \cite{BuchholzRoos}. Although based on a similar idea,
they developed a somewhat complementary framework.
Roughly speaking, while 
their theory covers   
the long distance aspects of local equilibrium,
we concentrate on the short distance
properties.
}

\vspace*{1cm}

\section*{Acknowledgments}
It is a pleasure to thank 
K.--H. Rehren for bringing
the artikel \cite{BuchholzRoos} 
to our attention, D. Bucholz for a discussion on
this artikel and L. Stuller 
for pointing out the B\'enard effect to us.

\end{document}